\documentclass[a4paper,conference]{IEEEtran}
\IEEEoverridecommandlockouts

\usepackage{graphicx}
\usepackage{tikz}
\usepackage{pgfplots}
\usepackage{xcolor}
\usepackage{color}
\usepackage{colortbl}
\usepackage{amsthm}
\usepackage{amsmath}
\usepackage{bbm}
\usepackage{amsfonts}
\usepackage{amssymb}
\usepackage{mathtools}
\usepackage{cite}
\usepackage[nolist]{acronym}
\usepackage{booktabs}
\usepackage{diagbox}		
\usepackage{upgreek}
\usepackage{bbm}
\usepackage{Files/bm}
\usepackage{Files/winsnotation}
\usepackage{Files/Symbols_OPL}
\usepackage{Files/LVcolours}
\usepackage[ruled,vlined,noresetcount]{algorithm2e}
\usepackage{setspace}
\usepackage{comment}
\usepackage{array}
\usepackage{bbold}
\usepackage{lipsum}

\newcommand{\popsize}{K}
\newcommand{\nslot}{N_{\mathrm{s}}}
\newcommand{\T}{T}
\newcommand{\Rsum}{R_{\mathrm{sum}}}
\newcommand{\Kcont}{K_{\mathrm{c}}}
\newcommand{\Nc}{C}
\newcommand{\load}{\mathsf{G}}
\newcommand{\expectation}{\mathbb{E}}
\newcommand{\ratebpr}{r}
\newcommand{\inforate}{R}
\newcommand{\nbch}{M}
\newcommand{\flag}{\mathsf{F}}
\newcommand{\alphairsa}{a}
\newcommand{\converse}{\mathbb{G}}

\mathtoolsset{showonlyrefs}

\makeatletter
\def\endthebibliography{%
  \def\@noitemerr{\@latex@warning{Empty `thebibliography' environment}}%
  \endlist
}
\makeatother
\IEEEoverridecommandlockouts

\theoremstyle{definition}
\newtheorem{remark}{Remark}
\newtheorem{theorem}{Theorem}
\newtheorem{lemma}{Lemma}
\newtheorem{example}{Example}

\pgfplotsset{compat=1.17}

\begin{document}

\title{Irregular Repetition Slotted ALOHA\\ in an Information-Theoretic Setting}

\author{%
  \IEEEauthorblockN{Enrico Paolini\authorrefmark{1}, Lorenzo Valentini\authorrefmark{1}, Velio Tralli\authorrefmark{2}, Marco Chiani\authorrefmark{1}}
  \IEEEauthorblockA{\authorrefmark{1}CNIT/WiLab, DEI, University of Bologna, Italy}
    \IEEEauthorblockA{\authorrefmark{2}CNIT/WiLab, DE, University of Ferrara, Italy\\
  	Email: \{e.paolini,lorenzo.valentini13, marco.chiani\}@unibo.it; velio.tralli@unife.it }
}

\maketitle 

\begin{acronym}
\small
\acro{ACK}{acknowledgement}
\acro{AWGN}{additive white Gaussian noise}
\acro{BAC}{binary adder channel}
\acro{BCH}{Bose–Chaudhuri–Hocquenghem}
\acro{BPR}{Bar-David, Plotnik, Rom}
\acro{BS}{base station}
\acro{CDF}{cumulative distribution function}
\acro{CRA}{coded random access}
\acro{CRC}{cyclic redundancy check}
\acro{CRDSA}{contention resolution diversity slotted ALOHA}
\acro{CSA}{coded slotted ALOHA}
\acro{eMBB}{enhanced mobile broad-band}
\acro{FER}{frame error rate}
\acro{IFSC}{intra-frame spatial coupling}
\acro{i.i.d.}{independent and identically distributed}
\acro{IRSA}{irregular repetition slotted ALOHA}
\acro{LDPC}{low-density parity-check}
\acro{LOS}{line of sight}
\acro{MAC}{medium access control}
\acro{MIMO}{multiple input multiple output}
\acro{ML}{maximum likelihood}
\acro{MMA}{massive multiple access}
\acro{mMTC}{massive machine-type communication}
\acro{MPR}{multi-packet reception}
\acro{MRC}{maximal ratio combining}
\acro{PAB}{payload aided based}
\acro{PDF}{probability density function}
\acro{PGF}{probability generating function}
\acro{PHY}{physical}
\acro{PLR}{packet loss rate}
\acro{PMF}{probability mass function}
\acro{PRCE}{perfect replica channel estimation}
\acro{QPSK}{quadrature phase-shift keying}
\acro{RF}{radio-frequency}
\acro{SC}{spatial coupling}
\acro{SIC}{successive interference cancellation}
\acro{SIS}{successive interference subtraction}
\acro{SNB}{squared norm based}
\acro{SNR}{signal-to-noise ratio}
\acro{TAC}{ternary adder channel}
\acro{URLLC}{ultra-reliable and low-latency communication}
\end{acronym}

\setcounter{page}{1}

\begin{abstract}
An information-theoretic approach to irregular repetition slotted ALOHA (IRSA) is proposed. In contrast with previous works, in which IRSA analysis is conducted only based on quantities that are typical of collision models such as the traffic, the new approach also captures more fundamental quantities. Specifically, a suitable codebook construction for the adder channel model is adopted to establish a link with successive interference cancellation over the multi-packet reception channel. This perspective allows proving achievability and converse results for the average sum rate of IRSA multiple access schemes.
\end{abstract}


\section{Introduction}

Recent years have witnessed a revival of random channel access theory and techniques, fostered by new applications such as grant-free access schemes in \ac{MMA} scenarios \cite{Chen2021:Massive}.
This renewed interest has brought back the problem, known for decades (e.g., \cite{wolf1981:coding,Gallager1985:Perspective}), of framing random access under an information-theoretic setting able to capture not only collisions and intermittent users' activity but also the physical communication process.
Several research efforts have been made into this direction over the years, among which we mention \cite{Mathys1990:Class,Bar-David1993:Forward,Luo2012:New,Chen2017:Capacity,Recep2021:Random}.
Moreover, many recent works in this area have considered an ``unsourced'' model \cite{Polyanskiy2017:Perspective}, in which all users have the same codebook, the receiver is only tasked with producing an unordered list of messages, and the main performance metric is the per-user error probability \cite{Ordentlich2017:low_complexity,Ngo2021:Random_user_activity,vem2019:unsourced,Fengler2021:SPARCs, Liva2021:Coding}.

Coded random access has recently emerged as a new approach to reliably support \ac{MMA} applications, featuring a combination of packet-based coding and \ac{SIC} and having a strong connection with codes on sparse graphs.
Instances of coded random access are \ac{CSA} \cite{paolini2015:csa} and \ac{IRSA} \cite{liva2011:irsa}; in the latter one, packet repetitions are used as a form of packet erasure coding.

Notably, so far the above-mentioned efforts towards a new framework for random access have addressed coded random access only marginally, which motivates us to explore this direction of investigation.
Therefore, in this paper we propose a new approach to \ac{IRSA}, capturing information-theoretic concepts such as codebook construction and information rate.
The specific choice of the users' codebooks, based on \cite{Bar-David1993:Forward}, and the adopted adder channel mode allows incorporating in the analysis some known key quantities, such as the \ac{IRSA} asymptotic load threshold over the \ac{MPR} channel. 
This allows us obtaining achievability and converse results on the average sum rate in the regime in which both the users' population size and the number of slots per frame grow unbounded, their ratio remaining constant.


\section{IRSA over MPR Channel}\label{sec:irsa_mpr}

\subsection{Irregular Repetition Slotted ALOHA}
\ac{IRSA} is a grant-free and uncoordinated access protocol for \ac{MMA}, strongly related to codes on sparse graphs \cite{liva2011:irsa}.
The protocol is grant-free as users perform transmissions with no prior handshake procedure with the receiver. 
It is uncoordinated as users cannot cooperate with each other.
The time is slotted and framed, and users are slot- and frame-synchronous. The number of slots per frame is $\nslot$, the total number of users is $\popsize$, the number of simultaneously contending users over the same frame, unknown to the receiver, is $\Kcont \leq \popsize$.

At the beginning of each frame some users are contending and some others are idle. 
Each non-idle user contends for transmission of one packet over the current frame.
The user samples, independently of the other users, a discrete random variable $D \in \{2,\dots,\bar{d}\}$ with probability distribution $\Pr(D=d) = \Lambda_d$ and \ac{PGF} 
$\Lambda(x) = \sum_{d} \Lambda_d x^d$. 
The distribution $\Lambda$ is the same for all users and is hereafter referred to as the \ac{IRSA} distribution.
The user then draws, uniformly at random and without replacement, $d$ integers between $0$ and $\nslot-1$ and transmits $d$ replicas of its packet in the corresponding slots of the frame.
The status of the frame after transmissions can be described by means of a bipartite graph with $\popsize$ user nodes, one per user, and $\nslot$ slot nodes, one per slot.
An edge connects user node $u_i$, $i\in\{0,\dots,\popsize-1\}$, to slot node $s_j$, $j\in\{0,\dots,\nslot-1\}$, if user $i$ performed transmission of a packet replica in slot $j$.

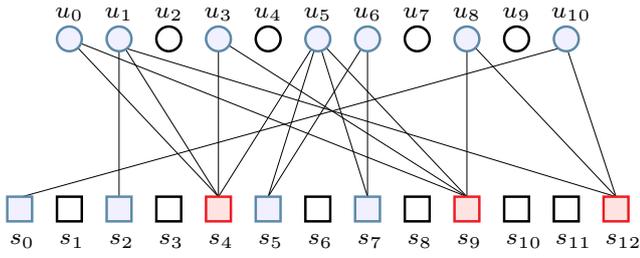
\begin{figure}[t]
\centering
         \resizebox{0.50\textwidth}{!}{%
     		\tikzset{every picture/.style={line width=0.75pt}} 

\begin{tikzpicture}[x=0.75pt,y=0.75pt,yscale=-1,xscale=1]

\draw [line width=0.25pt]  (18,66) -- (78,130);
\draw [line width=0.25pt]  (18,66) -- (178,130);

\draw [line width=0.25pt]  (38,66) -- (38,130);
\draw [line width=0.25pt]  (38,66) -- (78,130);
\draw [line width=0.25pt]  (38,69) -- (238,130);

\draw [line width=0.25pt]  (78,66) -- (78,130);
\draw [line width=0.25pt]  (78,66) -- (178,130);

\draw [line width=0.25pt]  (118,66) -- (78,130);
\draw [line width=0.25pt]  (118,66) -- (98,130);
\draw [line width=0.25pt]  (118,66) -- (138,130);
\draw [line width=0.25pt]  (118,66) -- (178,130);

\draw [line width=0.25pt]  (138,66) -- (98,130);
\draw [line width=0.25pt]  (138,66) -- (138,130);

\draw [line width=0.25pt]  (178,66) -- (178,130);
\draw [line width=0.25pt]  (178,66) -- (238,130);

\draw [line width=0.25pt]  (218,69) -- (-1,130);
\draw [line width=0.25pt]  (218,69) -- (238,130);

\draw (18,56) node  [font=\scriptsize] [align=left] {$\displaystyle u_{0}$};

\draw  [color={rgb, 255:red, 92; green, 138; blue, 168 }  ,draw opacity=1 ][fill={rgb, 255:red, 240; green, 240; blue, 255 }  ,fill opacity=1 ] (13,66) .. controls (13,63.24) and (15.24,61) .. (18,61) .. controls (20.76,61) and (23,63.24) .. (23,66) .. controls (23,68.76) and (20.76,71) .. (18,71) .. controls (15.24,71) and (13,68.76) .. (13,66) -- cycle ;

\draw (38,56) node  [font=\scriptsize] [align=left] {$\displaystyle u_{1}$};

\draw  [color={rgb, 255:red, 92; green, 138; blue, 168 }  ,draw opacity=1 ][fill={rgb, 255:red, 240; green, 240; blue, 255 }  ,fill opacity=1 ] (33,66) .. controls (33,63.24) and (35.24,61) .. (38,61) .. controls (40.76,61) and (43,63.24) .. (43,66) .. controls (43,68.76) and (40.76,71) .. (38,71) .. controls (35.24,71) and (33,68.76) .. (33,66) -- cycle ;

\draw (58,56) node  [font=\scriptsize] [align=left] {$\displaystyle u_{2}$};

\draw  (53,66) .. controls (53,63.24) and (55.24,61) .. (58,61) .. controls (60.76,61) and (63,63.24) .. (63,66) .. controls (63,68.76) and (60.76,71) .. (58,71) .. controls (55.24,71) and (53,68.76) .. (53,66) -- cycle ;

\draw (78,56) node  [font=\scriptsize] [align=left] {$\displaystyle u_{3}$};

\draw  [color={rgb, 255:red, 92; green, 138; blue, 168 }  ,draw opacity=1 ][fill={rgb, 255:red, 240; green, 240; blue, 255 }  ,fill opacity=1 ] (73,66) .. controls (73,63.24) and (75.24,61) .. (78,61) .. controls (80.76,61) and (83,63.24) .. (83,66) .. controls (83,68.76) and (80.76,71) .. (78,71) .. controls (75.24,71) and (73,68.76) .. (73,66) -- cycle ;

\draw (98,56) node  [font=\scriptsize] [align=left] {$\displaystyle u_{4}$};

\draw  (93,66) .. controls (93,63.24) and (95.24,61) .. (98,61) .. controls (100.76,61) and (103,63.24) .. (103,66) .. controls (103,68.76) and (100.76,71) .. (98,71) .. controls (95.24,71) and (93,68.76) .. (93,66) -- cycle ;

\draw (118,56) node  [font=\scriptsize] [align=left] {$\displaystyle u_{5}$};

\draw  [color={rgb, 255:red, 92; green, 138; blue, 168 }  ,draw opacity=1 ][fill={rgb, 255:red, 240; green, 240; blue, 255 }  ,fill opacity=1 ] (113,66) .. controls (113,63.24) and (115.24,61) .. (118,61) .. controls (120.76,61) and (123,63.24) .. (123,66) .. controls (123,68.76) and (120.76,71) .. (118,71) .. controls (115.24,71) and (113,68.76) .. (113,66) -- cycle ;

\draw (138,56) node  [font=\scriptsize] [align=left] {$\displaystyle u_{6}$};

\draw  [color={rgb, 255:red, 92; green, 138; blue, 168 }  ,draw opacity=1 ][fill={rgb, 255:red, 240; green, 240; blue, 255 }  ,fill opacity=1 ] (133,66) .. controls (133,63.24) and (135.24,61) .. (138,61) .. controls (140.76,61) and (143,63.24) .. (143,66) .. controls (143,68.76) and (140.76,71) .. (138,71) .. controls (135.24,71) and (133,68.76) .. (133,66) -- cycle ;

\draw (158,56) node  [font=\scriptsize] [align=left] {$\displaystyle u_{7}$};

\draw  (153,66) .. controls (153,63.24) and (155.24,61) .. (158,61) .. controls (160.76,61) and (163,63.24) .. (163,66) .. controls (163,68.76) and (160.76,71) .. (158,71) .. controls (155.24,71) and (153,68.76) .. (153,66) -- cycle ;

\draw (178,56) node  [font=\scriptsize] [align=left] {$\displaystyle u_{8}$};

\draw  [color={rgb, 255:red, 92; green, 138; blue, 168 }  ,draw opacity=1 ][fill={rgb, 255:red, 240; green, 240; blue, 255 }  ,fill opacity=1 ] (173,66) .. controls (173,63.24) and (175.24,61) .. (178,61) .. controls (180.76,61) and (183,63.24) .. (183,66) .. controls (183,68.76) and (180.76,71) .. (178,71) .. controls (175.24,71) and (173,68.76) .. (173,66) -- cycle ;

\draw (198,56) node  [font=\scriptsize] [align=left] {$\displaystyle u_{9}$};

\draw  (193,66) .. controls (193,63.24) and (195.24,61) .. (198,61) .. controls (200.76,61) and (203,63.24) .. (203,66) .. controls (203,68.76) and (200.76,71) .. (198,71) .. controls (195.24,71) and (193,68.76) .. (193,66) -- cycle ;

\draw (220,56) node  [font=\scriptsize] [align=left] {$\displaystyle u_{10}$};

\draw  [color={rgb, 255:red, 92; green, 138; blue, 168 }  ,draw opacity=1 ][fill={rgb, 255:red, 240; green, 240; blue, 255 }  ,fill opacity=1 ] (213,66) .. controls (213,63.24) and (215.24,61) .. (218,61) .. controls (220.76,61) and (223,63.24) .. (223,66) .. controls (223,68.76) and (220.76,71) .. (218,71) .. controls (215.24,71) and (213,68.76) .. (213,66) -- cycle ;

\draw [color={rgb, 255:red, 92; green, 138; blue, 168 }  ,draw opacity=1 ][fill={rgb, 255:red, 240; green, 240; blue, 255 }  ,fill opacity=1 ]  (-7,130) -- (3,130) -- (3,140) -- (-7,140) -- cycle ;
\draw (-1,148) node  [font=\scriptsize] [align=left] {$\displaystyle s_{0}$};

\draw   (13,130) -- (23,130) -- (23,140) -- (13,140) -- cycle ;
\draw (19,148) node  [font=\scriptsize] [align=left] {$\displaystyle s_{1}$};

\draw [color={rgb, 255:red, 92; green, 138; blue, 168 }  ,draw opacity=1 ][fill={rgb, 255:red, 240; green, 240; blue, 255 }  ,fill opacity=1 ]  (33,130) -- (43,130) -- (43,140) -- (33,140) -- cycle ;
\draw (39,148) node  [font=\scriptsize] [align=left] {$\displaystyle s_{2}$};

\draw   (53,130) -- (63,130) -- (63,140) -- (53,140) -- cycle ;
\draw (59,148) node  [font=\scriptsize] [align=left] {$\displaystyle s_{3}$};

\draw [color={rgb, 255:red, 237; green, 28; blue, 36 }  ,draw opacity=1 ][fill={rgb, 255:red, 255; green, 227; blue, 224 }  ,fill opacity=1 ]  (73,130) -- (83,130) -- (83,140) -- (73,140) -- cycle ;
\draw (79,148) node  [font=\scriptsize] [align=left] {$\displaystyle s_{4}$};

\draw  [color={rgb, 255:red, 92; green, 138; blue, 168 }  ,draw opacity=1 ][fill={rgb, 255:red, 240; green, 240; blue, 255 }  ,fill opacity=1 ] (93,130) -- (103,130) -- (103,140) -- (93,140) -- cycle ;
\draw (99,148) node  [font=\scriptsize] [align=left] {$\displaystyle s_{5}$};

\draw   (113,130) -- (123,130) -- (123,140) -- (113,140) -- cycle ;
\draw (119,148) node  [font=\scriptsize] [align=left] {$\displaystyle s_{6}$};

\draw  [color={rgb, 255:red, 92; green, 138; blue, 168 }  ,draw opacity=1 ][fill={rgb, 255:red, 240; green, 240; blue, 255 }  ,fill opacity=1 ] (133,130) -- (143,130) -- (143,140) -- (133,140) -- cycle ;
\draw (139,148) node  [font=\scriptsize] [align=left] {$\displaystyle s_{7}$};

\draw   (153,130) -- (163,130) -- (163,140) -- (153,140) -- cycle ;
\draw (159,148) node  [font=\scriptsize] [align=left] {$\displaystyle s_{8}$};

\draw [color={rgb, 255:red, 237; green, 28; blue, 36 }  ,draw opacity=1 ][fill={rgb, 255:red, 255; green, 227; blue, 224 }  ,fill opacity=1 ]  (173,130) -- (183,130) -- (183,140) -- (173,140) -- cycle ;
\draw (179,148) node  [font=\scriptsize] [align=left] {$\displaystyle s_{9}$};

\draw   (193,130) -- (203,130) -- (203,140) -- (193,140) -- cycle ;
\draw (201,148) node  [font=\scriptsize] [align=left] {$\displaystyle s_{10}$};

\draw   (213,130) -- (223,130) -- (223,140) -- (213,140) -- cycle ;
\draw (221,148) node  [font=\scriptsize] [align=left] {$\displaystyle s_{11}$};

\draw [color={rgb, 255:red, 237; green, 28; blue, 36 }  ,draw opacity=1 ][fill={rgb, 255:red, 255; green, 227; blue, 224 }  ,fill opacity=1 ]  (233,130) -- (243,130) -- (243,140) -- (233,140) -- cycle ;
\draw (241,148) node  [font=\scriptsize] [align=left] {$\displaystyle s_{12}$};

\end{tikzpicture}
     	}%
       \caption{Graphical representation of \ac{IRSA} access with $\popsize=11$ users ($\Kcont=7$ contending) and $\nslot=13$ slots. Light-blue circles are contending users and blank circles idle users. Assuming $\T=2$, light-blue squares are resolvable slots, light-red squares collision slots, and blank squares empty slots.}
    \label{fig:frame}
\end{figure}

\begin{example}
An example with $\popsize = 11$ users ($u_0$, $\dots$, $u_{10}$) and $\nslot = 13$ slots ($s_0$, $\dots$, $s_{12}$) is depicted in Fig.~\ref{fig:frame}.
Users are represented by circles and slots by squares.
There are $\Kcont = 7$ contending users, represented by colored circles; the number of transmitted packet replicas may change from user to user.
Blank circles represent users that are idle in the current frame.
\end{example}

The efficiency of \ac{IRSA}, denoted by $\eta$, is defined as $\eta = [\Lambda'(1)]^{-1}$ where $\Lambda'(1)$ represents the expected number of replicas transmitted by a contending user.
If users become contending independently of each other with probability $\pi$ at the beginning of each frame, the average load is 
$\load = \pi \popsize/\nslot\,\mathrm{[packets/slot]}
$.
The average number of packet replicas per slot is $\load / \eta$.

\subsection{\ac{SIC}-Based \ac{IRSA} Receiver over the \ac{MPR} Channel}

The $\T$-\ac{MPR} channel is a packet-based channel model representing a simple extension of the collision channel \cite{Ghez1988:Stability}.
The $\T$-\ac{MPR} channel model is based on three assumptions that may be summarized as follows:

\emph{Assumption~1}: The receiver can always discriminate between a slot with no packet arrivals, a slot with at most $\T$ arrivals, and a slot with more than $\T$ arrivals;
    
\emph{Assumption~2}: If there are more than $\T$ arrivals in a slot, a condition detected by the receiver, then none of these packets can be correctly received;
    
\emph{Assumption~3}: If there are at most $\T$ arrivals in a slot, a condition detected by the receiver, then all of these packets are correctly received with zero error probability.

Over the $\T$-\ac{MPR} channel model, the \ac{IRSA} receiver attempts recovery of all packets using an iterative \ac{SIC}-based procedure.
At the beginning, the receiver marks all slots as empty, resolvable (at most $\T$ arrivals), or collision (more than $\T$ arrivals) slots.
Then, each iteration comprises two steps and may be described as follows, under the assumption that each replica carries information about the number and the positions of the other replicas:

1. In every resolvable slot, all packets are correctly received.

2. For each such packet, the interference of every replica of the packet is subtracted from the corresponding slot, possibly leading to new resolvable slots.

\begin{example}
With reference again to Fig.~\ref{fig:frame}, assume $\T=2$. 
Since slots $s_0$, $s_2$, $s_5$, and $s_7$, are resolvable, packet replicas transmitted by users $u_1$, $u_5$, $u_6$, and $u_{10}$ in these slots are correctly received at the first iteration.
Interference cancellation in the first iteration makes slots $s_4$ and $s_{12}$ resolvable.
In the second iteration, at least one packet replica transmitted by users $u_0$, $u_3$, and $u_8$ is correctly received in these slots.
Interference cancellation in the second iteration makes all slots empty and the procedure terminates successfully.
\end{example}

The \ac{SIC}-based procedure has been analyzed in the asymptotic setting $\nslot \rightarrow \infty$, $\popsize \rightarrow \infty$, and $\alphairsa=\popsize/\nslot$ constant \cite{Ghanbarinejad2013:Irregular,stefanovic2018:multipacket}, under the following further assumption:
   
\emph{Assumption~4}: Interference subtraction across slots is ideal.

The analysis resembles density evolution of LDPC codes over the erasure channel and shows the existence of an asymptotic load threshold $\load^{*} = \load^{*}(\Lambda,\T)$.
Let $\nslot \rightarrow \infty$, $\popsize \rightarrow \infty$, and $\alphairsa=\popsize/\nslot$ be constant. 
In this asymptotic setting, let $p_{\ell}$ be the probability that the generic edge is connected to a slot node not yet resolvable at the end of iteration $\ell$.
Then, the evolution of $p_{\ell}$ through \ac{SIC} iterations is governed by the recursion
\begin{align}
  p_{\ell} &= 1 - \exp\left( - \load \Lambda'(p_{\ell-1}) \right) \sum_{k = 0}^{\T - 1} \frac{ 1 }{k!} \left( \load \Lambda'(p_{\ell-1}) \right)^{k}
\end{align}
with starting point $p_1 = 1-\exp(-\frac{\load}{\eta})\sum_{k=0}^{\T-1} \frac{1}{k!} (\frac{\load}{\eta})^k$. 
The load threshold is defined as
\begin{align}\label{eq:MPR_threshold}
    \load^{*} = \sup \{ \load > 0 : p_{\ell} \rightarrow 0 \text{ as } \ell \rightarrow \infty\} \, .
\end{align}
\begin{remark}\label{remark:plp}
The quantity $p_{\ell}^d$ represents the probability that the packet of a user, that transmitted $d$ packet replicas, has not yet been correctly received after $\ell$ \ac{SIC} iterations.
Thus the condition $p_{\ell} \rightarrow 0$, which holds for all $\load \leq \load^{*}(\Lambda,\T)$, corresponds to a vanishing packet loss probability in the asymptotic setting.
\end{remark}


\section{A New Perspective on \ac{IRSA}} \label{sec:abc}

The description of \ac{IRSA} provided in Section~\ref{sec:irsa_mpr} is the usually adopted one and is typical of collision resolution approaches: it focuses on bursty arrivals and traffic, while it ignores any information-theoretic quantity. 
In this section we formulate a new description in order to fill this gap.
We address the channel model at symbol level, the multiple access code, and the decoding function.
We propose a framework for the $\T$-\ac{MPR} channel, including all these aspects, which, on one hand, represents a simple way to realize Assumptions 1-4 above and, on the other hand, provides a connection between random access method and encoding functions and sum rates.

\subsubsection{Channel model} The channel model we employ is the noiseless adder channel \cite{Dyachkov1981:Coding}.
The channel has $\popsize$ inputs $X_1$, $\dots$, $X_{\popsize}$ and one output $Y$, where $X_i \in \{0,1\} \subset \mathbb{R}$ for all $i \in \{1,\dots,\popsize\}$, $Y \in \{0,\dots,\popsize\} \subset \mathbb{R}$, and
\begin{align}
    Y = \sum_{i=1}^{\popsize} X_i
\end{align}
the sum being over the reals.
Due to the binary input alphabets we refer to this channel as the \ac{BAC}.

\subsubsection{Codebook construction and encoding} To define the encoding function of each user we resort on the codebook construction proposed by \ac{BPR} in \cite{Bar-David1993:Forward}, where a coding technique to achieve zero error probability over the ``$\popsize$-choose-$\T$'' \ac{BAC} is proposed.
In this channel model there are $\popsize$ users, of which exactly $\T$ are contending.
The $\T$ binary codewords are transmitted over a \ac{BAC}; the decoder aims at decoding the $\T$ messages, with $\T$ known a priori, and at associating each message to one of the $\popsize$ users.

The \ac{BPR} construction starts from the parity-check matrix of a $\T$-error-correcting binary \ac{BCH} code of length $\nbch=2^m-1$, in the form
\begin{align}\label{eq:H}
    \mathbf{H} = \left[ 
    \begin{array}{ccccc}
        1 & \alpha & \alpha^2 & \dots & \alpha^{2^m-2} \\
        1 & \alpha^3 & \alpha^6 & \dots & \alpha^{3(2^m-2)} \\
        \vdots & \vdots & \vdots & \vdots & \vdots \\
        1 & \alpha^{2\T-1} & \alpha^{2(2\T-1)} & \dots & \alpha^{(2\T-1)(2^m-2)}
    \end{array}
    \right]
\end{align}
with $m\T$ binary rows (as the binary image of $\alpha$, a primitive element of $\mathrm{GF}(2^m)$, has length $m$) and $\nbch$ columns.
The columns are partitioned into $\popsize$ disjoint subsets, each of size $\frac{\nbch}{\popsize}$ (assuming $\popsize$ divides $\nbch$) and each subset is employed as the codebook of the corresponding user.\footnote{The assumption that $\popsize$ divides $\nslot$ is made throughout the paper. In case $\nbch$ is not a multiple of $\popsize$, one can simply re-define $\nbch$ as the largest multiple of $\popsize$ that is less than $2^m-1$, and assign $M/\popsize$ columns of $\mathbf{H}$ to each user.}
All users have the same information rate and the same codeword length $m\T$.
Note that no codeword can be shared by any two users, that all codebooks are nonlinear, and that no codebook contains the all-$0$ codeword.
Note also that, although we stick to \ac{BCH} codes as in \cite{Bar-David1993:Forward}, the codebook construction may in principle rely on other error correcting code families.

The \ac{BPR} construction achieves zero error probability over the $\popsize$-choose-$\T$ \ac{BAC}.
In fact, since by construction any $2\T$ columns of the parity-check matrix of a $\T$-error-correcting linear block code are linearly independent, any two different $\T$-tuples of codewords cannot have the same sum.
Hence, the set of $\T$ messages can be decoded with no error at the receiver and its message can be uniquely associated with a user.

Next we address the proposed \ac{IRSA} encoder. 
We refer to the described coding scheme as ``\ac{IRSA}-\ac{BPR}''.
\ac{BPR} $(m,\T,\popsize)$ codebooks are assigned to the $\popsize$ users, as described above.
At the beginning of a frame, each user has an idle message $W=0$ (meaning nothing to communicate) with probability $1-\pi$ and one message of information $W \in \{1,\dots,\frac{\nbch}{\popsize}\}$ with probability $\pi$, independently from user to user. 
Users with a message $W \neq 0$ are contending ones and the message of each contending user is uniformly distributed in $\{1,\dots,\frac{\nbch}{\popsize}\}$.
An idle user generates an idle sequence $X^N(0)$ of $N = (1+m\T) \nslot$ symbols, all equal to $0 \in \mathbb{R}$.
On the other hand, a contending user generates a binary codeword $X^N(W)$ of length $N$ symbols, in the following way.
The message $W$ is encoded into a codeword of length $m\T$ belonging to the user's \ac{BPR} codebook and one extra-symbol equal to $1 \in \mathbb{R}$ is appended to the \ac{BPR} codeword.
Then, the length-$N$ codeword is constructed as the concatenation of $d$ ``\ac{BPR} blocks'', each containing the \ac{BPR} codeword with the extra symbol $1$, and of $\nslot-d$ ``idle blocks'', each of length $1+m\T$ and containing the symbol $0 \in \mathbb{R}$ only.
Importantly, to reflect the grant-free and uncoordinated nature of the access scheme, the number of \ac{BPR} blocks in the codeword, $d$, and the positions of these $d$ blocks are functions solely of the random message $W$.
A way to generate the repetition degree $d$ and the positions of the $d$ \ac{BPR} blocks out of $W$ is discussed in the Appendix.

\begin{remark}
Each of the $\nslot$ blocks composing a codeword $X^N(W)$ corresponds to one slot in the description provided in Section~\ref{sec:irsa_mpr}.
As such, the codeword of a contending user corresponds to the entire frame.
This is consistent with the message $W$ being transmitted using the channel $N$ times.
The information rate of a contending user is $\inforate=\ratebpr/\nslot$ where
\begin{align}\label{eq:rate_bpr}
    \ratebpr = \frac{1}{1+ m \T} \log_2 \frac{\nbch}{\popsize}
\end{align}
is the \ac{BPR} encoder rate with the extra symbol $1$.
The average sum rate, $\Rsum$, is given by $\Rsum = \expectation[\Kcont]\, \inforate$.
We note also that the encoder may be seen as the concatenation of two component encoders: a nonlinear \ac{BPR} encoder and an \ac{IRSA} ``random access'' encoder.
\end{remark}

\begin{remark}
The IRSA-BPR construction provides different codebooks to different users, thus allowing the receiver to identify the users that are sending messages. 
A slightly simplified IRSA coding scheme may be also defined for the unsourced setting, where each user can use the entire codebook without any prior assignment, and the receiver is only able to produce an unordered list of messages.  The information rate for this unsourced construction can be easily derived, leading to an encoder rate expression as in (1), but without the term $\log_2K$, which appears as the cost of identifying the users that send messages. However, we should note that in the unsourced model the function of message association to users is left to upper layer protocols that spend at least $\log_2K$ bits  in the overhead of each message to accomplish this task.
\end{remark}

\subsubsection{Iterative Decoding} The vector $Y^N$ at the output of the \ac{BAC} is a sequence of $\nslot$ blocks, each of length $1+m\T$ symbols.
We say a contending user is active in any such block when a \ac{BPR} block is present in that position in the user's codeword.
Decoding is performed iteratively; in every iteration all non-zero blocks of $Y^N$ are processed and, for each of them, \ac{BPR} decoding is run if the number of active users, known owing to the extra-symbol $1$, does not exceed $\T$.
Each message $W$ decoded from a block can be associated with no error to a specific user; $W$ is re-encoded and the obtained length-$N$ codeword is subtracted from $Y^N$.
No action is taken in a block if the number of active users in it exceeds $\T$.
Decoding terminates when $Y^N = 0^N$ (success) or when $Y^N \neq 0^N$ but no further subtraction can be performed (premature stop).

Denoting the iterative decoding function by $\phi(\cdot)$, we have 
\begin{align}
\phi(Y^N) = \{ (W_{1},\dots,W_{q}), (i_1,\dots,i_q), \flag \}
\end{align}
where $(W_1,\dots,W_q)$ is an ordered message list, $(i_1,\dots,i_q)$ is the ordered list of the corresponding user's indexes, $\flag=0$ flags a success ($q=\Kcont$, complete message list), and $\flag=1$ flags a premature stop ($q<\Kcont$, incomplete message list).

For any random access coding scheme we distinguish between the decoding error probability (probability that the obtained message list is different from the actual one and/or some incorrect user-message association is done) and the per-user error probability (probability that the message of a user is not in the decoded message list and/or that the user is not associated with its message).


\section{Results}\label{sec:results}

We provide both achievability and converse results on the expected sum rate.
The achievability one applies to the \ac{IRSA}-\ac{BPR} scheme, while the converse has a more general validity.
We always assume that, in each frame, each user is contending with probability $\pi$, independently of the other users.

\begin{theorem}\label{th:achievability}
Consider an \ac{IRSA}-\ac{BPR} scheme over the \ac{BAC}, with distribution $\Lambda$ and parameters $\T$ and $m = \frac{1}{\epsilon} \log_2 \popsize$, $0<\epsilon<1$.
If 
\begin{align}\label{eq:achievability} 
  \Rsum \leq \frac{1-\epsilon}{\T} \, \load^{*}(\Lambda,\T)
\end{align}
then the per user error probability under iterative decoding tends to zero in the limit as $\nslot \rightarrow \infty$ and constant $\alphairsa=\popsize/\nslot$, where $\load^{*}(\Lambda,\T)$ is the load threshold of the \ac{IRSA} distribution $\Lambda$ over the $\T$-\ac{MPR} channel model, defined in \eqref{eq:MPR_threshold}.
\end{theorem}
\begin{IEEEproof}
In the considered setting, Assumptions 1-4 reviewed in Section~\ref{sec:irsa_mpr} are all satisfied.
In fact, due to the addition of the extra symbol to each \ac{BPR} codeword, the receiver has constantly perfect knowledge of the number of \ac{BPR} blocks in each slot (Assumption~1) and takes no action in any slot with more than $\T$ such blocks (Assumption~2).
Moreover, up to $\T$ \ac{BPR} blocks in the same slot can always be decoded with zero error probability (Assumption~3).
Finally, over the \ac{BAC} interference subtraction is perfect since the channel is noiseless and no channel estimation is required. As such, iterative decoding for \ac{IRSA}-\ac{BPR} over the \ac{BAC} has a one-to-one correspondence with \ac{SIC} over the \ac{MPR} channel (Section~\ref{sec:irsa_mpr}) and the per-user error probability coincides with the packet loss probability mentioned in Remark~\ref{remark:plp}.
Next, the average sum rate is
\begingroup
\allowdisplaybreaks
\begin{align}
    \Rsum &= \expectation[\Kcont] \frac{\ratebpr}{\nslot} = \frac{\load}{1+ \frac{\T}{\epsilon} \log_2 \popsize} \log_2 \frac{\popsize^{\frac{1}{\epsilon}} -1}{\popsize} \label{eq:rsum_exact}
\end{align}
\endgroup
where \eqref{eq:rsum_exact} follows from \eqref{eq:rate_bpr} and $m = \frac{1}{\epsilon} \log_2 \popsize$.
As $\nslot \rightarrow \infty$ for constant $\alphairsa$, the per-user error probability vanishes for all $\load \leq \load^{*}(\Lambda,\T)$ under \ac{SIC} decoding. 
Since
\begin{align}\label{eq:limit_K_eps}
\frac{\log_2 \frac{\popsize^{\frac{1}{\epsilon}} -1}{\popsize} }{1+ \frac{\T}{\epsilon} \log_2 \popsize} \rightarrow \frac{1-\epsilon}{\T}   
\end{align}
as $\popsize = \alphairsa \nslot \rightarrow \infty$, we obtain the statement.
\end{IEEEproof}

\begin{remark}
We note that the slot size (number of channel uses per slot), and then the ``packet'' size, is $\Theta(\log_2 \popsize)$.
We may interpret this as a consequence of the decoder task to perform association between decoded messages and contending users.
\end{remark}

We now address a converse result. We start by proving the following useful lemma.

\begin{lemma}\label{lemma:loeve}
Let $C$ be any positive random variable.
For constant $\alphairsa = \popsize/\nslot$, if 
$
\lim_{\substack{\nslot \rightarrow \infty}} \Pr(\Kcont > \Nc) = 0
$
then
\begin{align}\label{eq:G_ineq_conv}
    \load \leq \lim_{\nslot \rightarrow \infty} \frac{\expectation[\Nc]}{\nslot} \, .
\end{align}
\end{lemma}
\begin{IEEEproof}
The statement can be proved by applying the simple lower bound on the probability of the upper tail provided in Lo{\'e}ve \cite{loeve1977:Probability}, stating that for a real random variable $X$ with support $\mathcal{X}$ and a nonnegative and nondecreasing function $g(\cdot) : \mathbb{R} \mapsto \mathbb{R}$ we have
\begin{align}\label{eq:loeve}
    \Pr(X > a) \geq \frac{\expectation[g(X)]-g(a)}{\sup g(x)}
\end{align}
where $\sup g(x) = \sup \{g(x) : x \in \mathcal{X} \}$.
Applying \eqref{eq:loeve} with $X = \Kcont - \Nc$, $g(x)=\max\{0,x\}$, and $a=0$, we obtain
\begingroup
\allowdisplaybreaks
\begin{align}
\Pr(\Kcont > \Nc) &\geq \frac{\expectation[\max\{0,\Kcont-\Nc\}]}{\sup g(k_{\mathrm{c}}-c)} \\
&\geq \frac{\max\{0,\expectation[\Kcont]-\expectation[\Nc]\}}{\sup g(k_{\mathrm{c}}-c)} \label{eq:jensen}
\end{align}
\endgroup
where \eqref{eq:jensen} follows from Jensen's inequality since $g(x)$ is convex.
By hypothesis, for any $\epsilon>0$ there exists $\nslot(\epsilon)$ such that $\Pr(\Kcont > \Nc) \leq \epsilon$ for all $\nslot > \nslot(\epsilon)$.
It follows that, for all $\nslot > \nslot(\epsilon)$, we also have
\begin{align}
\frac{\max\{0,\expectation[\Kcont]-\expectation[\Nc]\}}{\sup g(k_{\mathrm{c}}-c)} \leq \epsilon
\end{align}
and then
\begingroup
\allowdisplaybreaks
\begin{align}
\max\{0,\expectation[\Kcont]-\expectation[\Nc]\} &\leq \epsilon \, \sup g(k_{\mathrm{c}}-c) \\
&\leq \epsilon\, \popsize \label{eq:eps_K}
\end{align}
\endgroup
where \eqref{eq:eps_K} is due to $k_{\mathrm{c}} \leq \popsize$ and $c > 0$. Inequality \eqref{eq:eps_K} implies
$\pi \popsize -\expectation[\Nc] \leq \epsilon \, \popsize$. Dividing both sides by $\nslot$ we obtain
$\load \leq \frac{\expectation[\Nc]}{\nslot} + \epsilon \, \alphairsa$.
Taking the limit $\nslot \rightarrow \infty$, $\epsilon$ becomes arbitrarily small; noting that $\alphairsa$ is constant (so $\epsilon \, \alphairsa$ vanishes) yields the statement.
\end{IEEEproof}

\medskip
We can use Lemma~\ref{lemma:loeve} to obtain a converse result valid for any framed and slotted access scheme over the \ac{BAC} (even cooperative or grant-based) with \ac{BPR} codebooks assigned to users and the slot size matching the \ac{BPR} codeword size.
We keep appending the extra symbol $1$ to the \ac{BPR} codeword, although not strictly necessary for the proof.

\begin{theorem}\label{th:converse}
Let \ac{BPR} codebooks be assigned to the $\popsize$ users, with $m = \frac{1}{\epsilon} \log_2 \popsize$, and the channel model be the \ac{BAC}.
Assume each contending user generates one \ac{BPR} codeword of length $n=1+m\T$ out of its message and transmits replicas of it in any number of slots, where the slot size is also $n$, according to any access protocol.
Length-$n$ all-zero words are transmitted by the user in the other slots.
Moreover, let $l_i$ be the number of \ac{BPR} codewords added in slot $i$ and 
\begin{align}\label{eq:ci}
\Nc_i = \left\{ 
\begin{array}{ll}
   l_i  & \text{if } l_i \leq \T \\
   \T  & \text{if } l_i > \T .
\end{array}
\right.
\end{align}
Let $P_e$ be the error probability under optimum decoding.
If $P_e \rightarrow 0$ as $\nslot \rightarrow \infty$ for constant $\alphairsa = \popsize / \nslot$, then
\begin{align}\label{eq:converse}
\Rsum \leq \frac{1-\epsilon}{\T} \, \converse
\end{align}
where $\converse$ is the sup of the set of all $\load$ fulfilling \eqref{eq:G_ineq_conv} and $\Nc = \sum_{i=0}^{\nslot-1} \Nc_i$ with $\Nc_i$ defined in \eqref{eq:ci}.
\end{theorem}
\begin{IEEEproof}
An optimum, idealized decoder is the one aware of which users are active in each slot and formulating the decoding problem over the \ac{BAC} as the solution of a linear system of equations in which the unknowns are the $\Kcont$ \ac{BPR} codewords.
The number of linearly independent equations available in slot $i$ is $C_i$ defined in \eqref{eq:ci}, so the total number of such equations is $C = \sum_{i=0}^{\nslot-1} C_i$.
If the error probability of this optimum decoder vanishes as $\nslot \rightarrow \infty$ and constant $a = \popsize/\nslot$, then the probability that the number of unknown exceeds the number of equations, $\Pr(\Kcont > C)$, also vanishes.
Lemma~\ref{lemma:loeve}, with \eqref{eq:rsum_exact} and \eqref{eq:limit_K_eps}, then leads to the statement. 
\end{IEEEproof}

\begin{remark}
Comparing \eqref{eq:achievability} and \eqref{eq:converse}, we can note how their structure is analogous.
As mentioned above, however, while Theorem~\ref{th:achievability} applies to \ac{IRSA}-\ac{BPR} under \ac{SIC} decoding, Theorem~\ref{th:converse} has a broader scope. 
A more subtle difference is that Theorem~\ref{th:achievability} considers the per-user error probability under iterative decoding while Theorem~\ref{th:converse} the decoding error probability under an optimum decoder.
\end{remark}

\begin{figure}
    \begin{center}
    \includegraphics[width=0.98\columnwidth]{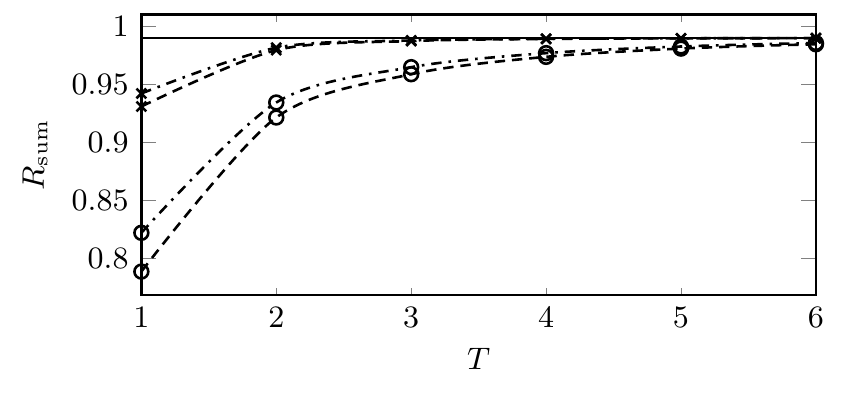}
    \end{center}
    \vspace{-5mm}
    \caption{Converse \eqref{eq:converse} versus $\T$ for $\epsilon=0.99$. Solid: coordinated scheme; dashed: \ac{BPR}-\ac{IRSA} scheme; dot-dashed: mixed scheme. \ac{IRSA} efficiency $\eta=1/2$ ($\mathsf{o}$) and $\eta=1/3$ ($\mathsf{x}$).}
    \label{fig:converse}
\end{figure}

\begin{example}
Consider the following three slotted and framed access schemes: (i) A coordinated scheme where each contending user fills one slot with a \ac{BPR} block and the other slots with all-zero blocks; the slot index is scheduled by the receiver to maximize the sum rate. (ii) An uncoordinated scheme where all contending users employ \ac{IRSA}-\ac{BPR}. (iii) A mixed scheme in which $\nu \popsize$ users have high priority and are served in a coordinated fashion, as in scheme (i), while the other $(1-\nu) \popsize$ users employ \ac{IRSA}-\ac{BPR}; $\gamma \nslot$ slots are reserved to high-priority users while the other $(1-\gamma) \nslot$ slots are available to low-priority ones, but the high-priority slots still empty after scheduled transmissions can be used by low-priority users as per decoder feedback.
Specifying \eqref{eq:G_ineq_conv} for the three schemes, we obtain $\lim_{\nslot \rightarrow \infty} \expectation[C]/\nslot = \min \{\load,T\}$, hence $\converse=\T$, for scheme (i). Proof sketch: $\Kcont/\nslot \rightarrow \load$ in probability as $\nslot \rightarrow \infty$; $f(x)=\min\{x,\T\}$ is continuous so $C/\nslot=f(\Kcont/\nslot) \rightarrow f(\load)$ in probability; $C/\nslot$ is bounded then we can conclude that $\expectation[C/\nslot] \rightarrow f(\load)$. For scheme (iii) we obtain (proof omitted)
%
%
\begin{align*}
    \load &\leq \T - (1 - \beta)
    \exp\! \bigg(\!\! - \frac{\load (1-\nu)}{\eta (1-\beta)} \bigg) \!\sum_{t=0}^{\T-1} \frac{\T-t}{t!} \bigg( \frac{\load (1-\nu)}{\eta (1 - \beta)} \bigg)^{\!t}
\end{align*}
where $\beta = \min \{ \frac{\nu \load}{\T}, \gamma \}$; this expression holds also for scheme (ii) with $\nu=\beta=0$. In both schemes (ii) and (iii), $\converse$ must be computed numerically.
The obtained values of \eqref{eq:converse} are plotted in Fig.~\ref{fig:converse} versus $\T$ for $\epsilon=0.99$.
Two values of $\eta$ are considered, namely, $\eta=1/2$ and $\eta=1/3$; concerning the mixed scheme, the plot is for $\nu=\gamma=0.2$.
The region above each curve is not achievable by the corresponding scheme.
The value of $\converse$ for scheme (ii) also appears in \cite{stefanovic2018:multipacket}, but it was proved only under \ac{SIC}, while it is shown here that it holds for any decoder.
\end{example}

\section{Conclusion}\label{sec:conclusions}

In this paper a new framework for \ac{IRSA} has been proposed, in which rigorous results on the average sum rate can be established.
A number of possible research directions, we believe, arise from this work. 
Among them, the development of converse results capturing the \ac{IRSA} distribution (and not just its efficiency), extension to \ac{CSA} schemes, extensions to channel models incorporating noise and fading and to different settings, e.g., to asynchronous schemes or to a finite number of users generating an increasing traffic as the frame size grows.


\section*{Appendix}
This appendix discusses generation of the repetition rate $d$ and of a $d$-tuple of different integers in $\{0,\dots,\nslot-1\}$ as a function of a message $W$ uniformly distributed in $\{1,\dots,\frac{\nbch}{\popsize}\}$.
For $\frac{\nbch}{\popsize} \Lambda_d$ an integer for all $d\in\{2,\dots,\bar{d}\}$, an instance of the random variable $D$ with \ac{PGF} $\Lambda(x)$, mentioned in Section~\ref{sec:irsa_mpr}, can be generated as
\begin{align}\label{eq:d_from_W}
    d = \min \bigg\{ i \in \{2,\dots,\bar{d}\} : \frac{\nbch}{\popsize} \sum_{h=2}^{i} \Lambda_h \geq W \bigg\} \, .
\end{align}
Once the repetition degree $d$ has been obtained from \eqref{eq:d_from_W}, the positions of the $d$ \ac{BPR} blocks can be generated by resorting on the combinadic representation of an integer (e.g., \cite{Knuth2011:Art}).
Accordingly, for any integer $0 \leq u < \binom{\nslot}{d}$, there exists a unique $d$-tuple of integers $(u_1,\dots,u_d)$, with $0 \leq u_1 < \dots < u_{d}$, such that 
%
$u = \binom{u_1}{1} + \binom{u_2}{2} + \dots + \binom{u_d}{d}$. %
In particular, for $h = d, \dots, 1$, $u_h$ is computed as
\begin{align}\label{eq:u_h}
    u_h = \max \bigg\{ i \in \mathbb{N} : \binom{i}{h} \leq u - \sum_{j=h+1}^{d} \binom{u_j}{j} \bigg\} \, .
\end{align}
The positions of the $d$ blocks can be computed by first letting $v = W-\frac{\nbch}{\popsize}\sum_{i=2}^{d-1} \Lambda_i-1$ (note that $v \in \{0,\dots,\frac{\nbch}{\popsize}\Lambda_d-1\}$), then by defining $u = v \mod \binom{\nslot}{d}$, and finally by returning $(u_1, \dots, u_d)$ through \eqref{eq:u_h}.
The desired uniform probability distribution over the $\binom{\nslot}{d}$ combinations is obtained when $\binom{\nslot}{d}$ divides $\frac{\nbch}{\popsize} \Lambda_d$ for all $d\in\{2,\dots,\bar{d}\}$.

\section*{Acknowledgement}
This work has been carried out in the framework of the CNIT National Laboratory WiLab and the WiLab-Huawei Joint Innovation Center.
The authors would like to thank Dr. Lei Wang for useful discussions.




\end{document}